\documentclass{article}
\pdfoutput=1
\usepackage{amsmath,graphicx,spconf}
\usepackage{mathtools}
\usepackage{amsfonts}
\usepackage{xcolor}
\usepackage{url}  
\usepackage{makecell}
\usepackage{cite}
\usepackage{balance}
\usepackage{tikz,fullpage}
\usepackage{booktabs}
\usepackage{multicol}
\usepackage{multirow}

\newcommand{\algalign}[2]

\newlength{\bibitemsep}
\newlength{\bibparskip}
\let\oldthebibliography\thebibliography
\renewcommand\thebibliography[1]{%
  \oldthebibliography{#1}%
  \setlength{\parskip}{\bibitemsep}%
  \setlength{\itemsep}{\bibparskip}%
}

\def\by{\mbox{$\mathbf y$}}

\def\bp{\mbox{$\mathbf p$}}

\def\bz{\mbox{$\mathbf z$}}
\def\by{\mbox{$\mathbf y$}}

\def\b1{\mbox{\boldmath $1$}}

\newcommand{\qedsymbol}{\hspace{\fill}\rule{1.5ex}{1.5ex}}

\title{Pooling Strategies for Simplicial Convolutional Networks }
\name{Domenico Mattia Cinque, Claudio Battiloro, Paolo Di Lorenzo}
\address{DIET Department, Sapienza University of Rome, Via Eudossiana 18, 00184, Rome, Italy \\
E-mail: domenico.cinque98@gmail.com, \{claudio.battiloro, paolo.dilorenzo\}@uniroma1.it 
}

\begin{document}
\usetikzlibrary{positioning,calc,arrows,shapes,decorations.pathmorphing,automata,backgrounds,petri}
\ninept

\maketitle

\begin{abstract}

\noindent The goal of this paper is to introduce pooling strategies for simplicial convolutional neural networks. Inspired by graph pooling methods, we introduce a general formulation for a simplicial pooling layer that performs: i) local aggregation of simplicial signals; ii) principled selection of sampling sets; iii) downsampling and simplicial topology adaptation. The general layer is then customized to design four different pooling strategies (i.e., max, top-$k$, self-attention, and separated top-$k$) grounded in the theory of topological signal processing. Also, we leverage the proposed layers in a hierarchical architecture that reduce complexity while representing data at different resolutions. Numerical results on real data benchmarks (i.e., flow and graph classification) illustrate the advantage of the proposed methods with respect to the state of the art.
\end{abstract}


\vspace{-.1cm}
\section{Introduction}
\label{sec:intro}

In the last years, Graph Neural Networks (GNNs) \cite{Bronstein2017GeometricDL, bruna2013spectral, kipf2016semi} have shown remarkable results in learning tasks involving data defined on irregular domains (e.g., graphs), such as social networks, recommender systems, cybersecurity, natural language processing, genomics, and many more \cite{emrgField, kipf2016semi}. However, GNNs are designed to work with graphs, which consider only pairwise relationships between data. On the contrary, many real-world phenomena involve multi-way relationships as, e.g., in biological or social networks. Some recent works in topological signal processing \cite{Barbarossa2020TopologicalSP,schaub2020random} have shown that multi-way relationships can be described using simplicial complexes, which are specific instances of hyper-graphs with powerful algebraic representation able to model higher-order interactions among nodes. Consequently, there was also a raising interest in the development of (deep) neural network architectures able to handle data defined on topological spaces, as summarized in the sequel.

\noindent \textbf{Related works.} Despite its recent birth, many contributions have been made to the field of simplicial deep learning. In \cite{snn}, the authors introduced a basic simplicial neural network (SNN) architecture that performs convolution exploiting high-order Laplacians without independently exploiting upper and lower neighbourhoods. In \cite{mpnn}, message passing neural networks (MPNNs)  are adapted to simplicial complexes, with the aggregation and updating functions taking into account data defined on adjacent simplices, enabling message exchange even among signals of different orders. The work in \cite{scn} exploits the simplicial filters introduced in \cite{firFilters} to design a flexible and low-complexity simplicial convolutional networks (SCNs) with spectral interpretability. Finally, in \cite{san,goh2022simplicial}, simplicial attentional architectures are introduced. 

Motivated by the fact that, both in convolutional neural networks (CNNs) and in GNNs, the introduction of pooling layers was proved to be useful for reducing the number of model parameters while improving the learning performance, in this work we aim to endow SCNs with pooling strategies. However, while for CNNs the pooling operation relies on aggregation based on the natural local neighbourhood provided by the regular grid domain, even on simpler graph domains the definition of local patches is not straightforward. Early works tried to overcome this issue by using graph clustering algorithms such as GraClus \cite{graclus} or spectral methods \cite{eigenpool} to produce a node assignment that generalizes the notion of locality present in regular domains. The most recent trends are instead focused on differentiable learnable operators that can learn a node assignment \cite{zhang2019hierarchical}, or simply keep some nodes while discarding the others \cite{Topk, lee2019self}. Other works \cite{cangea2018towards} discuss the class of \emph{global pooling} methods that reduce the graph to a single vector, ignoring topological information. 
To the best of our knowledge, no previous works tackled the problem of pooling for SCNs.

\noindent \textbf{Contribution.} The goal of this work is to introduce pooling strategies for SCNs. Taking inspiration from the select-reduce-connect (SRC) paradigm \cite{UnderstandPool}, we introduce a general simplicial pooling layer that comprises three steps: i) a local aggregation step responsible for providing a meaningful summary of the input signals;  ii) a selection step responsible for selecting a proper subset of simplices; finally, iii) a reduction step that downsamples the input complex and the aggregated signals of step i) based on the simplices selected in step ii). By tailoring steps ii) and iii), we introduce four different simplicial pooling layers that generalize the well-known graph pooling strategies. Also, we exploit the proposed simplicial pooling layers in a jumping knowledge (JK) hierarchical architecture \cite{xu2018jumpknow}, which aggregates the intermediate embeddings produced by the simplicial pooling layers to produce the final output. Finally, we assess the performance of the proposed methods on real-world graph and trajectory classification tasks, showing favorable comparisons with respect to other techniques in terms of performance and robustness to compression.

\vspace{-.2cm}
\section{Background}\label{sec:background}
\vspace{-.1cm}

\noindent \textbf{Simplicial complex and signals.} Given a finite set of vertices $\cal{V}$, a $k$-simplex $\mathcal{H}_k$ is a subset of $\cal{V}$ with cardinality $k+1$. A face of $\mathcal{H}_k$ is a subset with cardinality $k$ and thus a $k$-simplex has $k + 1$ faces. A coface of $\mathcal{H}_k$ is a $(k+1)$-simplex that includes $\mathcal{H}_k$ \cite{Barbarossa2020TopologicalSP, lim2020hodge}. The lower neighbourhood $\mathcal{N}_{\downarrow}$ of $\mathcal{H}_{k}$ is the set of simplices with which it shares a face. Similarly, the upper neighbourhood $\mathcal{N}_{\uparrow}$ of $\mathcal{H}_{k}$ is the set of simplices with which it shares a co-face. A simplicial complex $\mathcal{X}_K$ of order $K$ is a collection of $k$-simplices $\mathcal{H}_k$, $k=0,\dots,K$ such that, for any $\mathcal{H}_k\in\mathcal{X}_K$ then $\mathcal{H}_{k-1}\in\mathcal{X}_K$ if $\mathcal{H}_{k-1} \subset \mathcal{H}_k$. We denote the set of $k$-simplex in $\mathcal X_k$ as $\mathcal D_k:=\{\mathcal{H}_k : \mathcal{H}_k\in \mathcal X_K\}$, with $|\mathcal D_k|=N_k$ and $\mathcal D_k \subset \mathcal X_K$.

A $k$-simplicial signal is defined as a mapping from the set of all $k$-simplices contained in the complex to the real numbers:
\begin{equation}\label{simp_signal}
    \mathbf{z}_k:\mathcal D_k \rightarrow \mathbb R, \qquad k=0,1,\dots,K.
\end{equation} 
In this paper, w.l.o.g., we will focus on complexes $\mathcal X_2$ of order up to two, thus with a set of vertices $\mathcal D_0 = \cal V$ with $|\mathcal{V}|=V$, a set of edges $\mathcal D_1= \cal E$ with $|\mathcal{E}|=E$ and a set of triangles $\mathcal D_2=\cal T$  with $|\mathcal{T}|=T$. Given a simplex $\mathcal{H}_{k-1}\subset \mathcal{H}_{k}$, we write $\mathcal{H}_{k-1} \sim \mathcal{H}_{k} $ to indicate that the orientation of $\mathcal{H}_{k-1}$ is coherent with the one of $\mathcal{H}_{k}$, whereas $\mathcal{H}_{k-1} \not\sim \mathcal{H}_{k}$ if it is not.

\noindent \textbf{Algebraic representations.} The structure of a simplicial complex $\mathcal{X}_K$ is described by the set of its incidence matrices $\mathbf{B}_k$, with $k=1,\dots,K$. The $\mathbf{B}_k$'s describe which $k$-simplices are incident to which $(k-1)$-simplices: 
\begin{equation}\label{incidence}
    [\mathbf{B}_k]_{i,j}= 
    \begin{cases}
    0 & \mathcal{H}_{k-1, i}\not \subset \mathcal{H}_{k,j}; \\ 
    1 & \mathcal{H}_{k-1,i}\subset \mathcal{H}_{k,j}\text{ and } \mathcal{H}_{k-1,i}\sim\mathcal{H}_{k,j}; \\
    -1  & \mathcal{H}_{k-1,i}\subset \mathcal{H}_{k,j}\text{ and } \mathcal{H}_{k-1,i}\not \sim\mathcal{H}_{k,j}.
    \end{cases}
\end{equation}
From the incidence information, we can build the high order combinatorial Laplacian matrices \cite{Golberg} of order $k = 0, \dots , K$:
\begin{align} 
&\mathbf{L}_0 = \mathbf{B}_1 \mathbf{B}_1^T \nonumber \\
&\mathbf{L}_k = \mathbf{B}_k^T \mathbf{B}_k +\mathbf{B}_{k+1} \mathbf{B}_{k+1}^T  =\mathbf{L}_{k,d}+\mathbf{L}_{k,u} \label{Lk}\\
&\mathbf{L}_{k} = \mathbf{B}_{K}^T \mathbf{B}_{K} . \nonumber\end{align}
The term $\mathbf{L}_{k,d}$ in \eqref{Lk}, also known as lower Laplacian, encodes the lower adjacency of $k$-order simplices; the second term $\mathbf{L}_{k,u}$, also known as upper Laplacian, encodes the upper adjacency of $k$-order simplices. Thus, for example, two edges are lower adjacent if they share a common vertex, whereas they are upper adjacent if they are faces of a common triangle. 
Note that the vertices of a graph can only be upper adjacent if they are incident to the same edge. This is why $\mathbf{L}_0$ contains only one term, and it corresponds to the usual graph Laplacian.\\
\noindent \textbf{Hodge Decomposition.} High order Laplacians admit a Hodge decomposition \cite{lim2020hodge}, 
such that any $k$-simplicial signal $\mathbf{z}_k\in \mathbb R^{N_k}$ can be decomposed as:
\begin{equation}\label{HodgeDec}
    \bz_k = \mathbf{B}_k^T\bz_{k-1}+\mathbf{B}_{k+1}\bz_{k+1} + \bz_{k,h},
\end{equation}
for $k=0,1,\dots,K$. The first term $\mathbf{B}_k^T\bz_{k-1}$ of  \eqref{HodgeDec} is called \textit{irrotational} component, the second term $\mathbf{B}_{k+1}\bz_{k+1}$ \textit{solenoidal} component, and the third term $\bz_{k,h}$ \textit{harmonic} component. In the sequel, we will focus w.l.o.g. on edge flow signals and complexes of order 2. Therefore, we will drop the subscripts and denote $\mathbf{z}_1$ with $\mathbf{z}$, $\mathbf L_1$ with $\mathbf L$, $\mathbf L_{1,d}$ with $\mathbf L_{d}$, $\mathbf L_{1,u}$ with $\mathbf L_{u}$ and $\mathcal{X}_2$ with $\mathcal{X}$. Moreover, we denote the lower and upper neighborhoods of the $i$-th edge with $\mathcal{N}_{i,\downarrow}$ and $\mathcal{N}_{i,\uparrow}$, respectively.

\begin{figure*}[t] 
\centering
\resizebox{0.82\textwidth}{!}{%
\begin{tikzpicture}[bend angle=45,auto,x=0.4cm,y=0.4cm]

  \tikzstyle{vert}=[circle,fill, thick,inner sep=1.3pt]
  \tikzstyle{transition}=[rectangle,thick,draw=black!75,
  			  fill=black!20,minimum size=4mm]

  \begin{scope}
    \node [vert] (v1) at (0,0) {};
    \node [vert] (v2) at (2,0) {};
    \node [vert] (v3) at (0,-2) {};
    \node [vert] (v4) at (2,-2) {};
    \node [vert] (v5) at (0,-4) {};
    \node [vert] (v6) at (2,-4) {};
    \node [vert] (v7) at (1,-3) {};
    
    \begin{pgfonlayer}{background}
    \path[draw=black!60] (v1) 
    edge node {} (v2)
    edge node {} (v3)
    edge node {} (v4);
    \path[draw=black!60] (v4) 
    edge node {} (v2)
    edge node {} (v3)
    edge node {} (v7)
    edge node {} (v6);
    \path[draw=black!60] (v3) 
    edge node {} (v5)
    edge node {} (v7);
    \path[draw=black!60] (v5) 
    edge node {} (v7)
    edge node {} (v6);
    \path[draw=black!60] (v6) 
    edge node {} (v7);
    \end{pgfonlayer}

    \begin{pgfonlayer}{background}
     \draw[fill=gray!80, opacity=0.20] (v1.center) -- (v2.center) -- (v4.center) -- (v1.center);
     \draw[fill=gray!80, opacity=0.20] (v4.center) -- (v7.center) -- (v6.center) -- (v4.center);
     \draw[fill=gray!80, opacity=0.20] (v3.center) -- (v7.center) -- (v5.center) -- (v3.center);
      \draw[fill=gray!80, opacity=0.20] (v7.center) -- (v6.center) -- (v5.center) -- (v7.center);
     \end{pgfonlayer}

  \end{scope}
  
  \begin{scope}[xshift=3.5cm]
    \node [vert] (v1') at (0,0) {};
    \node [vert] (v2') at (2,0) {};
    \node [vert] (v3') at (0,-2) {};
    \node [vert] (v4') at (2,-2) {};
    \node [vert] (v5') at (0,-4) {};
    \node [vert] (v6') at (2,-4) {};
    \node [vert] (v7') at (1,-3) {};

    \begin{pgfonlayer}{background}
    \path[draw=black!60] (v1') 
    edge node {} (v2')
    edge node {} (v4');
    
    \path[draw=red!40, style=dashed] (v1') 
    edge node {} (v3');
    
    \path[draw=black!60] (v4') 
    edge node {} (v2')
    edge node {} (v7')
    edge node {} (v6');
    \path[draw=black!60] (v3') 
    edge node {} (v7');
    
    \path[draw=red!40, style=dashed] (v3') 
    edge node {} (v4')
    edge node {} (v5');
    
    \path[draw=black!60] (v5') 
    edge node {} (v7');
    
    \path[draw=red!40, style=dashed] (v5') 
    edge node {} (v6');

    \path[draw=black!60] (v6') 
    edge node {} (v7');
    \end{pgfonlayer}
    
    \begin{pgfonlayer}{background}
     \draw[fill=gray!80, opacity=0.20] (v1'.center) -- (v2'.center) -- (v4'.center) -- (v1'.center);
     \draw[fill=gray!80, opacity=0.20] (v4'.center) -- (v7'.center) -- (v6'.center) -- (v4'.center);
     \end{pgfonlayer}
  \end{scope}
  
  \begin{scope}[xshift=7.0cm]
    \node [vert] (v1'') at (0,0) {};
    \node [vert] (v2'') at (2,0) {};
    \node [vert] (v3'') at (0,-2) {};
    \node [vert] (v4'') at (2,-2) {};
    \node [vert] (v5'') at (0,-4) {};
    \node [vert] (v6'') at (2,-4) {};
    \node [vert] (v7'') at (1,-3) {};

    \begin{pgfonlayer}{background}
    \path[draw=black!60] (v1'') 
    edge node {} (v4'');
    
    \path[draw=red!40, style=dashed] (v1'') 
    edge node {} (v2'');
        
    \path[draw=black!60] (v4'') 
    edge node {} (v2'')
    edge node {} (v7'')
    edge node {} (v6'');
    
    \path[draw=red!40, style=dashed] (v3'') 
    edge node {} (v7'');

    \path[draw=black!60] (v5'') 
    edge node {} (v7'');

    \path[draw=black!60] (v6'') 
    edge node {} (v7'');
    \end{pgfonlayer}
    
    \begin{pgfonlayer}{background}
     \draw[fill=gray!80, opacity=0.20] (v4''.center) -- (v7''.center) -- (v6''.center) -- (v4''.center);
     \end{pgfonlayer}
  \end{scope}

  \begin{scope}[xshift=10cm]
    \node [shape=circle, draw, thick, inner sep=1.5pt] (sum) at (0,-2) {$\sum$};
  \end{scope}

    \draw [-to, thick, decorate, decoration={snake, amplitude=.4mm,
         segment length=2mm}]
    ([xshift=3mm] v4 -| v6) -- ([xshift=-3mm]v3'-| v5')
    node [below=1mm,midway,text centered, inner sep=1pt]
      {\textsc{Pool}} 
    node [above=1mm,midway,text centered, inner sep=1pt]
      {\textsc{SCN}};
      
    \draw[-to, thick, decorate, decoration={snake, amplitude=.4mm,
         segment length=2mm}]
    ([xshift=3mm] v4' -| v6') -- ([xshift=-3mm]v3''-| v5'')
    node [below=1mm,midway,text centered, inner sep=1pt]
      {\textsc{Pool}} 
    node [above=1mm,midway,text centered, inner sep=1pt]
      {\textsc{SCN}};
      
    \draw [-to,thick, decorate, decoration={snake, amplitude=.4mm,
         segment length=2mm}, out=-0,in=-140]
    ([yshift=-2mm, xshift=2mm] v5' -| v6') to (sum) 
    node at (v6'.east)[xshift=7mm, yshift=-6mm]
      {\text{readout}};
    
    \draw [-to,thick, decorate, decoration={snake, amplitude=.4mm,
         segment length=2mm}, out=0,in=140]
    ([xshift=3mm] v1''-| v2'') to (sum) 
    node at (v2''.east) [xshift=10mm, yshift=3mm]
      {\text{readout}};
      
    \draw [-to,thick,decorate, decoration={snake, amplitude=.4mm,
         segment length=2mm}]
    (sum) to ([xshift=20mm] sum) 
    node at (sum) [xshift=11mm, yshift=3.5mm]
      {\textsc{MLP}};
 
\end{tikzpicture}
}
\caption{Example of simplicial pooling and JK hierarchical architecture.}
\label{fig:layer}
\end{figure*}
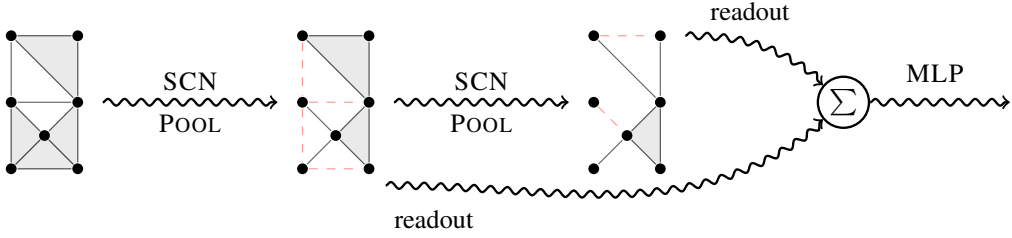

\section{Simplicial Convolutional Networks with Pooling Layers}
The Hodge decomposition in \eqref{HodgeDec} suggests to separately filter the components of simplicial signals. Indeed, the work in \cite{firFilters} introduced linear shift-invariant (LSI) filters for simplicial signals, which can be seen as a generalization of LSI graph filters that exploit both upper and lower connectivities. A simplicial convolutional neural network is made by the concatenation of several layers composed by a point-wise non-linearity applied to a bank of LSI simplicial filters plus a residual connection \cite{scn}. In this paper, we generalize the layer structure of \cite{scn} introducing a a family of pooling strategies encoded into the mapping $\mathcal{P}(\cdot)$. In particular, letting $\mathbf{X} \in \mathbb{R}^{E \times G}$ be the matrix collecting $G$ edge signals on its columns, the layer of an SCN endowed with pooling mechanisms (SCNP) can be written as:
\vspace{-.1cm}
\begin{align} \label{sclayer}
\mathbf{Y}=\sigma \Bigg[\mathcal{P}\Bigg( &\underbrace{\sum_{p = 1}^{J_{d}}\mathbf{L}_{d}^p\mathbf{X}\mathbf{D}_{p}}_{\mathbf{Z}_d} +\underbrace{\sum_{p = 1}^{J_{u}} \mathbf{L}_{u}^p\mathbf{X}\mathbf{U}_{p}}_{\mathbf{Z}_u}  + \underbrace{\mathbf{X}\mathbf{H}}_{\mathbf{Z}_h}\Bigg) \Bigg],
\vspace{-.1cm}
\end{align}
where $\mathbf{Y} \in \mathbb{R}^{E' \times F}$, with $E' \leq E$; the filters and residual weights $\big\{\mathbf{D}_{p}\big\}_{p=1}^{J_{d}}$, $\big\{\mathbf{U}_{p}\big\}_{p=1}^{J_{u}}$ and $\mathbf{H}$ $\in \mathbb{R}^{G \times F}$ are learnable parameters, while the order $J_{d}$ and $J_{u}$ of the filters, the number $F$ of output signals, and the non-linearity $\sigma(\cdot)$ are hyperparameters to be chosen at each layer. Therefore, an SCNP of depth $L$ is built as the stack of $L$ layers defined as in \eqref{sclayer}; the SCN layer in \cite{scn} is recovered from \eqref{sclayer} removing the pooling stage.

\subsection{Design of simplicial pooling mapping}\label{sec:pooling}


In this paragraph, we present a general formulation for a simplicial pooling mapping, which will then be tailored to design four different pooling strategies. Let us first denote the input to the pooling mapping in (\ref{sclayer}) as  $\mathbf{Z}=\mathbf{Z}_d+\mathbf{Z}_u+\mathbf{Z}_h \in \mathbb{R}^{E \times F}$, and let the simplicial complex structure be encoded into  $\mathcal{X}=(\mathbf{L}_u, \mathbf{L}_d)$. We also denote as  $\mathbf{Z}' \in \mathbb{R}^{E' \times F}$ the output of the pooling layer in (\ref{sclayer}). Then, formally, we define a simplicial pooling layer as the mapping
\begin{equation}
\mathcal{P}:(\mathcal{X}, \mathbf{Z}) \mapsto (\mathcal{X}', \mathbf{Z}'),\label{pooling_op}
\end{equation} 
which takes as input a simplicial complex $\mathcal{X}$ and  signals $\mathbf{Z} \in \mathbb{R}^{E \times F}$ defined on it,
and returns as output a sub-complex $\mathcal{X}'\subset\mathcal{X}$ and signals $\mathbf{Z}'\in \mathbb{R}^{E' \times F}$ defined on it, with $E'< E$.

Following the pooling paradigm introduced in \cite{UnderstandPool} for GNNs, we propose to model the layer in \eqref{pooling_op} as the composition of three operations: a local aggregation step, a selection step, and a reduction step. The local aggregation step is responsible for providing  summary signals $\widetilde{\mathbf{Z}} \in \mathbb{R}^{E \times F}$ of the input signals $\mathbf{Z} \in \mathbb{R}^{E \times F}$ leveraging the connectivity induced by the complex $\mathcal{X}$. Formally, we define it as the local mapping: 
\begin{equation}
\textbf{(Aggregation)}\quad
\mathcal{A}
:(\mathcal{X}, \mathbf{Z}) \mapsto (\mathcal{X}, \widetilde{\mathbf{Z}}).    \label{agg_op}
\end{equation} 
The mapping in (\ref{agg_op}) is local in the sense that the aggregated signals $[\widetilde{\mathbf{Z}}]_i$ of the $i$-th edge are function only of the signals of its (lower and/or upper) neighbours $[\widetilde{\mathbf{Z}}]_j$, $j \in \mathcal{N}_{i,\downarrow}$ and/or $\mathcal{N}_{i,\uparrow}$.

The selection step is responsible for choosing a subset $\mathcal{E}' \subset \mathcal{E}$ of edges that will compose the 1-skeleton of the sub-complex $\mathcal{X}'$. Formally, we define it as a mapping:
\begin{equation}
\textbf{(Selection)}\quad
\mathcal{S}:(\mathcal{X}, \widetilde{\mathbf{Z}}) \mapsto \mathcal{E}' \label{select_op}.
\end{equation} 
The cardinality of $\mathcal{E}'$ is tuned via the pooling ratio  $r \in (0,1]$ (a hyperparameter to be chosen), such that $|\mathcal{E}'| = E' = \lfloor r \cdot E \rfloor$.

Finally, the reduction step is responsible for properly downsampling the input complex $\mathcal{X}$ and the aggregated signals $\widetilde{\mathbf{Z}}$ to obtain the output sub-complex $\mathcal{X}'$ and the output signals $\mathbf{Z}'\in \mathbb{R}^{E'\times F}$, based on the edge set $\mathcal{E}'$ chosen through the selection set. Formally, we define it as a mapping:
\begin{equation}
\textbf{(Reduction)}\quad
\mathcal{R}:(\mathcal{E}', \mathcal{X}, \widetilde{\mathbf{Z}}) \mapsto (\mathcal{X}', \mathbf{Z}') \label{reduce_op}.
\end{equation} 
We assume that the reduction mapping in (\ref{reduce_op}) is given by the concurrent application of two independent operations, i.e., $\mathcal{R}=(\mathcal{R}_{S},\mathcal{R}_{C})$, which separately downsample signal and simplicial complex structure, respectively, and are defined as:
\begin{align}
\textit{(Signal reduction)}\quad &\mathcal{R}_{S}:(\mathcal{E}', \widetilde{\mathbf{Z}}) \mapsto \mathbf{Z}', \label{reducesig_op} \\
\textit{(Complex reduction)}\quad &\mathcal{R}_{C}:(\mathcal{E}', \mathcal{X}) \mapsto \mathcal{X}'. \label{reducecom_op} 
\end{align} 
The operations \eqref{reducesig_op}-\eqref{reducecom_op} compute the signals $\mathbf{Z}'$ and the complex structure $\mathcal{X}'$ at the output of the pooling layer, respectively. 

In summary, the general pooling mapping $\mathcal{P}$ in \eqref{pooling_op} is given by the composition of the three operations in (\ref{agg_op})-(\ref{reduce_op}), i.e.,
\begin{equation}
\mathcal{P}= \mathcal{R} \circ \mathcal{S} \circ \mathcal{A}.\label{pooling_comp}
\end{equation} 
We assume that the aggregation in \eqref{agg_op} is kept fixed (e.g., max or mean). Also, the complex reduction in \eqref{reducecom_op} is computed as follows: if an edge $e$ belongs to $\mathcal{E}$ but it is not in $ \mathcal{E}'$, the lower connectivity is updated by disconnecting the nodes that are on the boundary of $e$, while the upper connectivity is updated by removing the triangles that have $e$ on their boundaries. 

\vspace{-.1cm}
\subsection{Simplicial pooling strategies}

In this paragraph, we customize the selection and signal reduction steps in \eqref{select_op} and \eqref{reducesig_op} to design four pooling strategies. 

\vspace{.1cm}
\noindent\textbf{Max pooling:} The first method is an extension of the Max Pooling strategy commonly used in CNNs. It selects the subset of edges $\mathcal{E}'$ by ranking the absolute values of the sum of the aggregated signals $\widetilde{\mathbf{Z}}$ of each edge. Formally, we define: 
\begin{align}
&\mathcal{S}: \quad \by = \left|\widetilde{\mathbf{Z}}\,\mathbf{1}\right|, \quad \mathcal{E}' = \textrm{top}_{E'}(\by), \\ 
&\mathcal{R}_{S}: \quad \mathbf{Z}' = [\widetilde{\mathbf{Z}}]_{i \in \mathcal{E}'},
\end{align} 
where $\mathbf{1}$ is the vector of all ones, and $\textrm{top}_{E'}(\cdot)$ selects the indexes of the  the $E'$ largest entries of its vector argument.

\vspace{.1cm}
\noindent\textbf{Top-$k$ pooling:} The next layer is a generalization of those proposed in \cite{Topk, cangea2018towards} for GNNs. It selects the subset of edges $\mathcal{E}'$ ranking a learnable weighted combination of the aggregated signals $\widetilde{\mathbf{Z}}$ of each edge. Then, it computes the reduced signals as a scaled version of $\widetilde{\mathbf{Z}}$ with coefficients in $[0,1]$ given by a normalization of the aforementioned weighted combination. Formally, we have:
\vspace{-.2cm}
\begin{align}
&\mathcal{S}: \quad \by = \frac{\widetilde{\mathbf{Z}}\,\bp}{\| \bp \|_2}, \quad \mathcal{E}' = \textrm{top}_{E'}(\by) \label{sel_topk}
\\ 
&\mathcal{R}_S: \quad \mathbf{Z}' = [\widetilde{\mathbf{Z}} \odot \tanh (\by \mathbf{1}^T)]_{i \in \mathcal{E}'}, \label{red_topk}
\vspace{-.1cm}
\end{align} 
where $\bp$ is a learnable vector, and $\odot$ is the Hadamard product. 
\begin{table*}[t]
    \centering
    \resizebox{2\columnwidth}{!}{%
    \begin{tabular}{llcccccc} 
    	\toprule
		& \multirow{2.5}{*}{Method} & \multicolumn{4}{c}{Graph Classification} & \multicolumn{2}{c}{Edge Flow Classification}  \\
			\cmidrule(lr){3-6} \cmidrule(lr){7-8}
      &  &  DD & PROTEINS & MSRC21 & NCI109 & Ocean Drifters & Synthetic Flow \\
      \midrule
\multirow{2}{*}{GCNs} &Top-$k$ & $78.82\pm3.49$ & $73.75\pm3.32$ & $85.61\pm5.60$ &  $74.35\pm2.76$ & N/A & N/A\\
&SelfAtt      & $78.52\pm1.88$ & $73.75\pm3.32$ & $89.12\pm10.33$ & $76.76\pm2.21$ & N/A & N/A\\
      \midrule
\multirow{6}{*}{SCNPs} & NoPool (SCNs \cite{scn}) & $78.57\pm4.64$ & $75.00\pm2.19$ & $92.98\pm2.48$  & $76.28\pm3.97$ & $98.25\pm1.11$ & $100.0\pm0.0$ \\
&Random & $84.37\pm6.68$ & $74.29\pm2.48$ & $93.33\pm3.80$ & $74.47\pm3.79$  & $76.25\pm6.72$ & $98.82\pm0.58$ \\
&Max            & $\mathbf{88.07}\pm3.86$ & $ 78.57\pm2.09$ & $\mathbf{96.14}\pm3.14$  & $72.92\pm4.60$& $100.0\pm0.0$ & $100.0\pm0.0$ \\
&Top-$k$           & $84.87\pm3.61$ & $79.29\pm2.31$ & $95.79\pm2.94$ &  $75.36\pm6.95$ & $\mathbf{100.0}\pm0.0$ & $100.0\pm0.0$ \\
&SelfAtt        & $86.39\pm4.38$ & $\mathbf{79.46}\pm4.14$ & $95.79\pm2.94$ &  $71.62\pm6.38$ & $86.13\pm5.89$ & $99.99\pm0.04$ \\
&SepTop-$k$        & $83.36\pm7.09$ & $74.64\pm4.22$ & $\mathbf{96.14}\pm0.78$  & $\mathbf{80.09}\pm2.35$ & $99.62\pm0.56$ & $\mathbf{100.0}\pm0.0$ \\
    \bottomrule
    \end{tabular}
}
     \caption{Accuracy on graph and trajectory classification}
     \label{table:results2}
\end{table*}

\vspace{.1cm}
\noindent\textbf{Self-Attention Pooling:} This method is a generalization of SagPool \cite{lee2019self}. The main difference with Top-$k$ is that the ranking is computed over the output of a simplicial convolutional layer as in \eqref{sclayer} without pooling and with one output signal, here briefly denoted as $\textsc{scn}$. Formally, we have:
\vspace{-.1cm}
\begin{align}
    &\mathcal{S}: \quad \by = \textsc{scn}(\widetilde{\mathbf{Z}},\mathbf{L}_{d},\mathbf{L}_{u}), \quad \mathcal{E}' = \textrm{top}_{E'}(\by), \\
    &\mathcal{R}_S: \quad \mathbf{Z}' = [\widetilde{\mathbf{Z}} \odot \tanh (\by \mathbf{ 1}^T)]_{i \in \mathcal{E}'}.
\end{align}

\vspace{.1cm}
\noindent\textbf{Separated Top-$k$ pooling:} The Hodge Decomposition in \eqref{HodgeDec} and the consequent structure of the SCNP layer in \eqref{sclayer} suggest to design pooling layers based on the computation of three different aggregated signals: $\widetilde{\mathbf{Z}}_d$ (obtained from $\mathbf{Z}_d$), $\widetilde{\mathbf{Z}}_u$ (obtained from $\mathbf{Z}_u$), and $\widetilde{\mathbf{Z}}_h$ (obtained from $\mathbf{Z}_h$). Consequently, we will have three corresponding score vectors $\by_{d}$, $\by_{u}$, and $\by_{h}$, respectively. Thus, the ``separated'' version of the Top$-k$ layer in \eqref{sel_topk}-\eqref{red_topk} is given by: 
\begin{align} &\mathcal{S}:\begin{dcases}
    \by_{d} = \frac{\widetilde{\mathbf{Z}}_d\, \bp_d}{\| \bp_d \|}, \; \by_u = \frac{\widetilde{\mathbf{Z}}_u\, \bp_u}{\| \bp_u \|}, \; \by_h = \frac{\widetilde{\mathbf{Z}}_h\, \bp_h}{\| \bp_h \|}\\ 
    \mathcal{E}' = \textrm{top}_{E'}(\by_d+\by_u+\by_h) \end{dcases}\\
&\mathcal{R}_S: \begin{dcases}
    &\mathbf{Z}'_d=[\widetilde{\mathbf{Z}}_d \odot \tanh (\by_d\mathbf{1}^T)]_{i \in \mathcal{E}'}  \\
    &\mathbf{Z}'_u=[\widetilde{\mathbf{Z}}_u \odot \tanh (\by_u\mathbf{1}^T)]_{i \in \mathcal{E}'}  \\
    &\mathbf{Z}'_h=[\widetilde{\mathbf{Z}}_h \odot \tanh (\by_h\mathbf{1}^T)]_{i \in \mathcal{E}'}  \\
    &\mathbf{Z}'= \mathbf{Z}'_d+  \mathbf{Z}'_u + \mathbf{Z}'_h  \end{dcases}
\end{align}
where $\bp_u$, $\bp_d$, and $\bp_h$ are learnable vectors. Also all the previous methods can be reformulated in this ``separated'' version, but we leave their presentation and assessment for future works.

\subsection{Hierarchical Architecture}\label{sec:hier}

In this paragraph, we introduce a JK hierarchical architecture aimed at exploiting the different data representations obtained after each pooling stage $l \in \{1,...,L\}$.  A pictorial overview of the proposed JK hierarchical architecture is shown in Fig. \ref{fig:layer}. In particular, applying a readout operation, each intermediate compact representation obtained at the output of layer $l$ collapses the current signals (and complex) into a single embedding vector. For instance, a possible choice is concatenating the mean and the maximum of the current signals.
These vectors are then aggregated to compose a global final embedding. For instance, if the same number of output signals is used at each layer, the intermediate representations can be summed to obtain a single global embedding vector. 
Finally, the global embedding can be passed through a multi-layer perceptron (MLP), if it is needed for the task. In the case of transductive (semisupervised) tasks, both the intermediate and global embeddings might be unnecessary and can be neglected. 

\vspace{-.2cm}
\section{Numerical Experiments}
\vspace{-.2cm}

In this section, we assess the performance of the proposed simplicial pooling layers and hierarchical architecture on two learning tasks: trajectory classification \cite{mpnn}, and real-world graph \cite{tud} classification \footnote{\scriptsize \url{https://github.com/domenicocinque/spm}}. We compare the four proposed simplicial pooling layers with a random pooling strategy, and with plain SCNs having no pooling layers. Also, for graph classification, we show the results obtained using GCNs \cite{kipf2016semi} equipped with the ``graph counterpart'' of the proposed simplicial pooling layers. All the hyper-parameters are tuned to obtain the best performance per each dataset, except for the pooling ratio, which we keep fixed at $r=0.7$. We compute the intermediate and global embeddings via mean-maximum concatenation and sum, respectively. The models are trained for 150 epochs using the Adam optimizer \cite{kingma2015adam} and early stopping with patience $25$. All the experiments are averaged on five random seeds.

\begin{figure}[t]
  \centering
  \hspace{-.3cm}\includegraphics[ width=0.8\columnwidth]{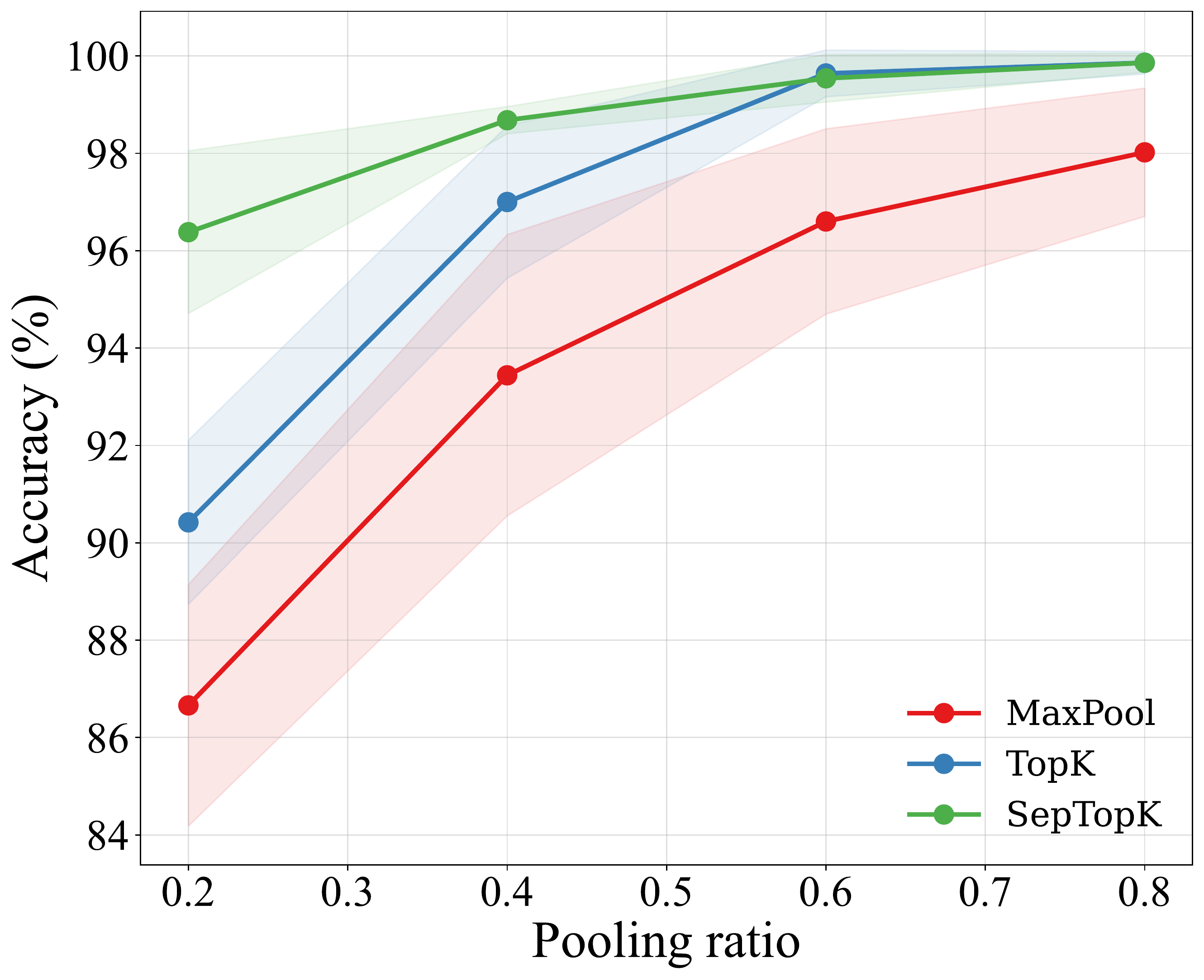}
  \vspace{-.2cm}
  \caption{Accuracy  versus  pooling ratio.}\label{acc_vs_ratio}\vspace{-.3cm}
\end{figure}

We first test the proposed simplicial pooling on two flow classification tasks, namely the synthetic flow and ocean drifter datasets, whose details can be found in  \cite{schaub2020random, mpnn}. In Table \ref{table:results2} (right side), we compare the accuracy obtained by all the considered methods, illustrating the gain introduced by the proposed simplicial pooling layers for both datasets. Then, to assess the accuracy-complexity tradeoff obtained by the proposed strategies, in Fig. \ref{acc_vs_ratio} we show the accuracy of the classification task versus the pooling ratio for the synthetic flow dataset considering three pooling methods. As we can see from Fig. \ref{acc_vs_ratio}, the accuracy mildly decreases with the pooling ratio, especially for the Separated Top-$k$ strategy, illustrating the very good accuracy-complexity trade-off obtained by the proposed methods. 

Finally, we study the performance of the proposed simplicial pooling layers on real-world graph classification tasks on the popular TUDataset \cite{tud} collection. To obtain simplicial complexes from the graphs, we follow the clique complex lifting procedure proposed in \cite{mpnn}, while the input edge signals are computed as the average of the graph signals of the boundary nodes. Then, in Table \ref{table:results2} (left side), we can see how the proposed SCNPs outperform SCNs with random pooling layers or no pooling, and the GCNs counterpart architectures.

\vspace{-.2cm}
\section{Conclusions}
\vspace{-.2cm}

In this paper, we have proposed a general formulation of a pooling layer for simplicial convolutional neural networks, designed as the composition of a local aggregation mapping, a selection mapping, and a reduction mapping. The proposed methodology is then tailored to design four different simplicial pooling layers, which generalize known graph pooling strategies for simplicial neural architectures. Numerical results on real and synthetic benchmarks illustrate the favorable performance of the proposed strategies with respect to other methods available in the literature. Future extensions include more complex simplicial architectures \cite{mpnn}, or cell complex neural networks \cite{bodnar2021weisfeiler, hajij2020cell, can}.

\clearpage

\balance
\bibliographystyle{IEEEbib}
\bibliography{refs}

\appendix

\end{document}